\documentclass[letterpaper, 10 pt, conference]{ieeeconf}  

\IEEEoverridecommandlockouts                              

\usepackage[l2tabu,orthodox]{nag}

\usepackage[
backend=biber, 
bibencoding=utf8,
style=ieee, 
sorting=none, 
natbib=true, 
doi=false, 
isbn=false, 
url=false, 
eprint=false, 
maxcitenames=1, 
mincitenames=1,
maxbibnames=5,
]{biblatex}

\usepackage[pdftex,colorlinks]{hyperref}

\usepackage[printonlyused]{acronym}
\acrodef{GNSS}{Global Navigation Satellite System}
\acrodef{RTK}{Real Time Kinematic}
\acrodef{ECU}{Electronic Control Unit}
\acrodef{MPC}{Model Predictive Control}
\acrodef{PID}{Proportional-Integral-Derivative}
\acrodef{IMU}{Inertial Motion Unit}
\acrodef{ANR}{French National Research Agency}
\acrodef{TIARA}{Toward Intelligent Adaptable Robots for Agriculture}
\acrodef{TSCF}{Technologies et systèmes d’information pour les agrosystèmes-Clermont-Ferrand}

\usepackage{siunitx}
\usepackage[all]{nowidow}

\usepackage[dvipsnames]{xcolor}

\usepackage{lipsum}

\usepackage[pdftex]{graphicx}
\usepackage{subcaption}
\usepackage{epstopdf}

\usepackage{import}

\graphicspath{{./latexGoodPractices/}}

\usepackage[dvipsnames]{xcolor}
\usepackage{tikz}
\usetikzlibrary{arrows, arrows.meta, calc, quotes,angles, patterns, intersections}

\usepackage{booktabs}

\usepackage{tabu}
\usepackage{soul}

\usepackage{amssymb,amsfonts,amsmath,amscd}

\usepackage{bm} 

\newcommand{\bbm}{\begin{bmatrix}}
\newcommand{\ebm}{\end{bmatrix}}

\usepackage{xspace}
\newcommand{\eg}{e.g.,\xspace}
\newcommand{\ie}{i.e.,\xspace}

\usepackage{microtype}
\usepackage{physics}

\title{\LARGE \bf
A Novel Control Strategy for Offset Points Tracking in the Context of Agricultural Robotics
}

\author{Stephane Ngnepiepaye Wembe$^{1,2}$, Vincent Rousseau$^{1,2}$, Johann Laconte$^{1}$ and Roland Lenain$^{1}$
\thanks{$^{1}$Université Clermont Auvergne, INRAE, UR TSCF, 63000, Clermont-Ferrand, France; stephane.ngnepiepaye-wembe@inrae.fr}%
\thanks{$^{2}$SABI AGRI, 63360, Saint-Beauzire, France; stephane.ngnepiepaye-wembe@sabi-agri.com}%
}

\begin{document}

\maketitle
\thispagestyle{empty}
\pagestyle{empty}

\begin{abstract}
In this paper, we present a novel method to control a rigidly connected location on the vehicle, such as a point on the implement in case of agricultural tasks.
Agricultural robots are transforming modern farming by enabling precise and efficient operations, replacing humans in arduous tasks while reducing the use of chemicals. 
Traditionally, path-following algorithms are designed to guide the vehicle’s center along a predefined trajectory. 
However, since the actual agronomic task is performed by the implement, it is essential to control a specific point on the implement itself rather than the vehicle’s center. 
As such, we present in this paper two approaches for achieving the control of an offset point on the robot. 
The first approach adapts existing control laws, initially intended for the rear axle’s midpoint, to manage the desired lateral deviation. 
The second approach employs backstepping control techniques to create a control law that directly targets the implement. 
We conduct real-world experiments, highlighting the limitations of traditional approaches for offset point control, and demonstrating the strengths and weaknesses of the proposed methods.
\end{abstract}

\section{Introduction}
The agricultural sector is faced with numerous challenges, including the need to increase production efficiency while simultaneously reducing environmental impacts. 
Traditional farming practices most often depend on manual labor and chemical inputs, which can result in inefficiencies and environmental contamination.
In response, the integration of robotics into agriculture is seen as a promising solution~\cite{lenain2021agricultural}. 
Agricultural robots offer the ability to perform repetitive tasks with high precision, such as planting, weeding, and harvesting, while reducing the need for chemical inputs, minimizing environmental footprints, and avoiding the use of manpower.
This growing interest is expressed in an array of research projects aimed at developing robotic systems for field operations, with the goal of enhancing both productivity and sustainability in agriculture~\cite{inbook_Robotics_in_Agriculture}.

\begin{figure}[t]
  \centering
  \includegraphics[width=\linewidth]{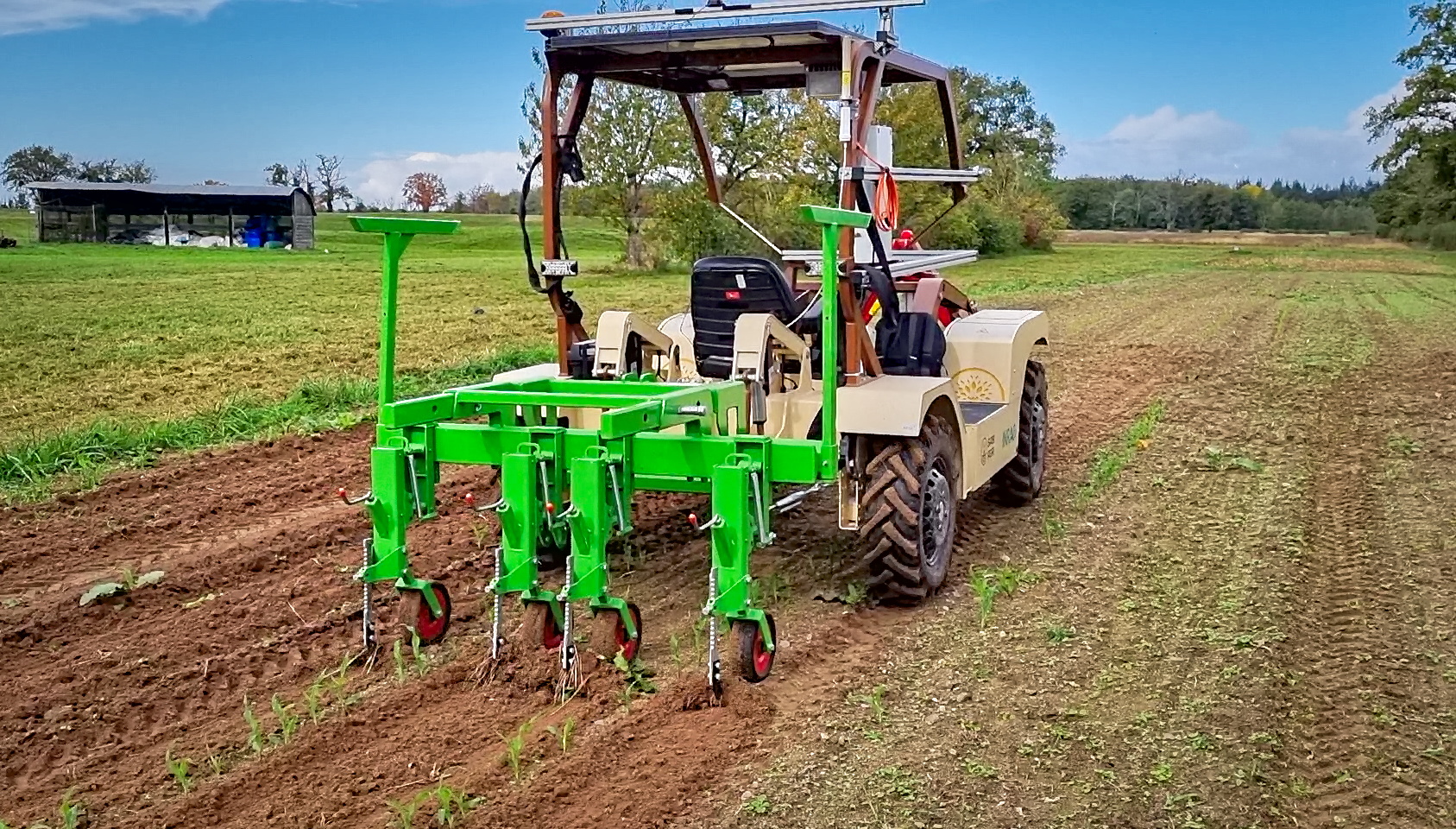}
  \caption{Agricultural robot equipped with a mounted implement. In this situation, the main focus is on the implement rather than the robot itself, as precision is required at the point where the implement interacts with the ground, thus making standard control laws of the robot's center inadequate.}
  \vspace{-1em}
  \label{fig:intro}
\end{figure}

A key challenge in agricultural robotics is controlling robots with implements.
Traditionally, control strategies in mobile robotics focus on the vehicle's kinematics, typically controlling the midpoint of the rear axle, which has convenient properties such as allowing an exact linearization of the kinematic model, thus advantageous for control~\cite{samson2016modeling}.
However, as depicted in \autoref{fig:intro}, in agricultural applications, the implement is the component that directly interacts with the soil and crops.
This difference requires a shift in the control techniques, from the robot to the implement itself. 
Current control approaches often deal with trailer-type implements, which are limited to pivoting configurations and attached at the rear of the vehicle.
However, many agricultural applications, such as using intercepts in viticulture or harrows in arable farming, require the implements to be mounted at the front of the vehicle or use rigid connections.
Thus, solely controlling the vehicle without considering the implement will degrade the overall precision of the agricultural task, from uneven seeding to damaging the crops in a weeding task~\cite{gan2007implement}.

As such, we propose in this paper a novel control algorithm that is specially designed to control an offset point of interest.
The method focuses on implements that are rigidly linked to the robot, as it is the case of most agricultural equipments.
Our contributions are threefold: 
1) A direct extension of the classical control method for controlling any offset point; 
2) A novel method specially designed for offset points, based on a backstepping approach; and
3) An evaluation on real-world experiments of the impact of the longitudinal distance of the implement over the path following error.

\section{Related Work}
The control of car-like wheeled vehicles has been extensively studied over the years, addressing various scenarios such as point-to-point movement, where the robot must travel from a known departure point to a specified destination~\cite{article_point_to_point}, trajectory tracking, which involves following a given path with a time constraint~\cite{inproceedings_Time_constrained_trajectory_tracking}, and path following, where the robot must match a specified trajectory without any time constraints~\cite{path_following_based_on_position}. 
The primary objective in these studies is to ensure that the robot's center of rear axle follows the desired trajectory accurately. 
Numerous control strategies have been developed to achieve this, including ``pure pursuit'' control laws~\cite{Coulter-1992-13338}, which define a target point on the reference trajectory at each instant and direct the robot towards this point~\cite{pure_pursuit}. 
While widely used, such a method can induce oscillatory behaviors. 
Alternative approaches, such as approximate linearization around the reference trajectory~\cite{feedback_linearisation} or exact linearization of the kinematic model~\cite{Samson_chained}, allow the application of methods from classical control theory, including PID controllers, to improve the system stability. 
Additionally, optimal control methods like Model Predictive Control (MPC) \cite{article_PMC} use predictive models to optimize control actions over a given time horizon, enhancing the system's performance. 
Backstepping methods \cite{Hadi_2020} remain a popular technique for controlling nonlinear systems, using a cascading approach where the control of the desired variable is achieved through intermediate variables, such as the angular deviation.

In the above works, the focus is on controlling the robot's center of inertia or the midpoint of the rear axle.
However, the complete control model for any point on the vehicle is highly nonlinear, which complicates the development of control laws. 
Moreover, the kinematic equation in any point but the rear axle cannot be exactly linearized through a variable transformation, making approaches such as \cite{akhtar_path_2011} unusable.
Consequently, the preferred option is to control the midpoint of the rear axle~\cite{lenain:hal-02603672}. 
However, in certain applications, the point of interest is not located at the vehicle's center, but at a specific point relative to it. 
In agricultural contexts, the implements attached to the vehicle perform the actual tasks, making the implement's position the primary point of interest rather than the midpoint of the vehicle's rear axle.

In the literature, two approaches are used to manage on-board implements. 
The first involves equipping the implement with its own control system and actuation, enabling it to follow the reference trajectory. 
\citet{semichev_automated_2020} propose the use of an automatic hitching that compensates the lateral deviation of the implement. 
\citet{freimann_basic_2007} proposes to integrate an \ac{ECU} making the implement and the tractor two distinct systems. 
Nevertheless, this implies the robot to have an active implement, which is not always the case.
This approach, which consists of suspending the implement and the tractor as two different systems, has its limitations, as disturbances from one system on the other are not taken into account. 
The second approach in the literature considers the implement and tractor as a single system, with the point of interest being a point of the implement. 
This approach is the most widespread in the literature, which propose methods for controlling a trailer~\cite{ zhe_leng_simple_2010} or a towed agricultural implement~\cite{article_fuzzy}. 
However, the connection between the implement and the vehicle is a pivot connection, allowing the trailer to pivot in curves, which helps to eliminate the lever arm effect and therefore the lateral error. 
In contrast, a rigidly attached implement will lead to increasing error as it is placed farther from the robot, as shown by \citet{gan2007implement}.
\citet{gartley2008online} provide an analysis of the change of dynamics while using an implement at the rear of the robot, while still controlling the center of the robot.
As such, to the best of our knowledge, the control of an offset point on the robot has not been studied.
Moreover, the cited approaches are specific to the case where the implement to be controlled is attached to the trailer at the rear of the robot, whereas multiple agricultural tasks are performed with a tool at the front of the robot.
As such, this paper presents a novel method to control any rigidly connected point on the robot, either at the front or back.

\section{Preliminaries}
In this section, we present ways to control an offset point (\eg an agricultural implement) on the robot.
First, the assumptions and modeling of the robot are described.
Then, the immediate extension of the classical control law is established, with a discussion on its limitations.
Finally, we show how to directly control the point of interesting via a backstepping approach.

\subsection{Modeling} 
In the following, we make the following assumptions:
\begin{description}
	\item[H1] All the vehicle wheels are in contact with the ground.
    \item[H2] The slippage and skidding effect are negligible.
	\item[H3] The dynamic effects are negligible.
    \item[H4] The offset point is linked via a rigid connection to the vehicle.
    \item[H5] The offset of the implement point must not exceed the minimum curvature radius of the trajectory.
	\item[H6] The robot has a vertical sagittal plane of symmetry passing through the center of the rear axle.
\end{description} 
Hypothesis H1 to H3 are verified when the vehicle is evolving at low speed and the robot's tires provide enough grip on the ground.
With these hypotheses, a kinematic model of the robot is enough to characterize its behavior.
Moreover, the less ideal case of slipping and skidding is left for future works.
Hypothesis H4 simply corresponds to the type of linkage agricultural vehicles possesses.
In that context, the implement is rigidly fixed to the vehicle, without any pivot link.
Agricultural machines typically use a three-point hitch to attach the implement, which is a rigid connection.
Hypothesis H5 is needed for the modeling described below to avoid undefined behaviors, as further discussed in the next part.
Finally, hypothesis H6 is verified for any ackerman-type vehicle, allowing to simplify the modeling of the robot to a bicycle-type vehicle. 

\begin{figure}[htbt]
    \centering
    \includegraphics[width=\linewidth]{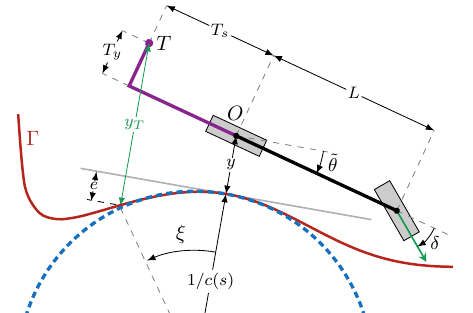}
    \caption{Notations used in the paper. The robot is assumed to follow a bicycle model. Contrary to standard robotics control problem, the goal is not to make the center $O$ of the robot to converge to the trajectory, but rather an offset point $T$ that represents in our case the implement (tool) of an agricultural robot. 
    }
    \label{fig:model}
\end{figure}

\autoref{fig:model} depicts the notations of the modeling of the robot.
As described above, hypothesis H6 allows one to model the robot using a bicycle model.
The offset point to control (tool) is denoted $T$, and can for instance represent the point of interaction between the implement and the ground.
As such, contrary to classical approaches, the center of the rear axle $O$ is not taken into account in the control law, and rather the offset point $T$ is considered.

The control point on the implement is defined by the distances $T_s$ and $T_y$, measured relative to the robot frame centered at the point $O$. 
Note that $T_s$ and $T_y$ can be either positive or negative, thus placing the the tool on the front, back, left or right of the robot's center.
In the following, $\tilde{\theta}$ denotes the angular deviation, and $y$ represents the tracking error at the center of the rear axle relative to the desired trajectory $\Gamma$, both measured using a \ac{GNSS} sensor and a geo-referenced trajectory.
The trajectory $\Gamma$ is assumed to be locally circular, with a curvature $c(s)$ at the curvilinear abscissa $s$. 
In this paper, we focus on regulating the implement lateral deviation $y_T$, defined as the deviation from the trajectory at the point on the implement closest to the trajectory, measured along a direction parallel to $y$.

Thus, the lateral deviation of the offset point $T$ is defined as
\begin{equation}
	\begin{aligned}
		y_T &= y+ T_s \sin\tilde{\theta} + T_y\cos{\tilde{\theta}}+ e,
	\end{aligned}
    \label{eq:RelationEcart}
\end{equation}
with
\begin{equation}
	\begin{aligned}
		e &= -\frac{1}{c(s)} \left( 1- \cos\xi \right), \quad\text{and}\\
		\xi &= \arcsin \left(c(s)(T_s\cos\tilde{\theta}+ T_y\sin\tilde{\theta})\right).
	\end{aligned}
\end{equation}
In the case where the curvature $c(s)$ is null, the distance $e$ is also null, as easily seen from \autoref{fig:model}.
Furthermore, \autoref{eq:RelationEcart} is well-defined if the quantity $c(s)(T_s\cos\tilde{\theta}+ T_y\sin\tilde{\theta})$ remains within the definition domain of the $\arcsin$ function, that is if 
\begin{equation}
    \left|T_s\cos\tilde{\theta} + T_y\sin\tilde{\theta}\right|<\left|\frac{1}{c(s)}\right|.
\end{equation} 
As this inequality must hold for all angular deviation $\tilde\theta$, the left-hand side attains its maximum at $\tilde\theta=\arctan{(T_s/T_y)}$, and thus we have the condition
\begin{equation}
\begin{aligned}
   \sqrt{T_s^2+T_y^2} < \left|\frac{1}{c(s)}\right|.
\end{aligned}
\label{eq:cond_TsTy}
\end{equation}
As such, the distance of the control point $T$ must not exceed the radius of curvature, and this corresponds to the hypothesis H5.
One can easily see that \autoref{eq:RelationEcart} is indeed undefined if the implement distance surpasses the curvature, as no projection of the tool position $T$ can be defined on the local osculating circle.

\subsection{Kinematic Model of the offset Point T}
In order to directly regulate the position of the tool $T$, it is necessary to have a kinematic model of the state variables.
Using the previous definitions, and the modeling described in~\cite{Lenain2004AdaptiveAP}, the kinematic of the lateral deviation with respect to time is written as
\begin{equation}
    \begin{aligned}
		\dot{y}_T & = v\sin\tilde{\theta} + \dot{\tilde{\theta}} \left(T_s  \cos\tilde{\theta} - T_y \sin\tilde{\theta} + \dv{e}{\tilde{\theta}}\right),
	\end{aligned} 
 \label{eq:DotYT}
\end{equation}
where $\dot{y}_T$ denoted the derivative with respect to time, and
\begin{equation}
    \dv{e}{\tilde{\theta}} = -c(s)\frac{(T_s \cos\tilde{\theta} + T_y \sin\tilde{\theta})(T_y \cos\tilde{\theta} - T_s \sin\tilde{\theta})}{\sqrt{1-c(s)^2(T_s \cos\tilde{\theta} + T_y \sin\tilde{\theta})^2}}.
    \label{eq:dedt}
\end{equation}
The above expression exists if $|T_s\cos\tilde{\theta} + T_y\sin\tilde{\theta}|<|\frac{1}{c(s)}|$, which is the same condition as \autoref{eq:cond_TsTy}.
As the term defined in \autoref{eq:dedt} is pre-multiplied by the curvature $c(s)$, it becomes negligible in regard to the other terms of \autoref{eq:DotYT} as long as the curvature remains small.
As a result, this term is assumed to be zero in the following, since heavy-duty agricultural machines have a limited maximum steering angle, and thus will only follow trajectories with small curvatures.
Future works will investigate the impact of this term in the case of larger curvatures.

Finally, according to~\cite{Lenain2004AdaptiveAP} and neglecting slip effects (hypothesis H2), the kinematics of the vehicle is given by
\begin{equation}
    \begin{aligned}
        \dot{s} &= \frac{v\cos{\tilde{\theta}}}{1-c(s)y}, \text{and}\\
        \dot{\tilde{\theta}}& = v\left(\frac{\tan\delta}{L} - \frac{c(s)\cos\tilde{\theta}}{1-c(s)y} \right), 
    \end{aligned} 
    \label{eq:DotTheata}
\end{equation}
where $v$ is the vehicle velocity.
Both quantities are defined if $y\neq 1/c(s)$, meaning that the position of the robot at the center of the osculating circle leads to singularities.
However, as the curvature is not reaching high values for most applications because of the robot's maximum steering angle, such an event is unlikely to happen.


\section{Control of the Offset Point $T$}
Using the kinematics of the control point defined above, we show how to derive two strategies to control an offset point on the robot.
First, a simple, direct refinement of the standard control algorithms is presented, followed by a discussion on its limitations.
Then, a more complex control law is proposed to answer these downsides.

\subsection{Control of a Desired Deviation of the Point $O$} \label{directe_control_section}
A straightforward and intuitive approach to control the tool's position along a trajectory is to apply a desired deviation at the point $O$, therefore guiding the tool $T$ along the path.
This method requires computing a desired lateral deviation $y^d$ to be applied for the tracking error $y$, allowing the convergence of the tool error $y_T$ to zero. 
As illustrated in~\autoref{fig:model}, tracking the point $O$ on the trajectory $\Gamma$ results in an offset for the point $T$:
 by incorporating this offset into the deviation, the point $T$ will be aligned with the trajectory. 

Assuming that the control law accurately regulates the lateral offset of the point $O$ to the desired deviation, and considering that the angular deviation is well regulated to zero ($\tilde{\theta}=0$), the necessary offset for the point $T$ to follow the reference trajectory can be expressed as
\begin{equation}
\begin{aligned}
		y^d & = -T_y - e^d,
\end{aligned} 
\label{eq:DefYD}
\end{equation}
with $e^d$ defined as
\begin{equation}
	 \begin{aligned}
		e^d & = -\frac{1}{c(s)} \left( 1- \cos\xi^d \right), \quad\text{and}\\
		\xi^d & = \arcsin \left(T_sc(s) \right).
	\end{aligned}
 \label{eq:DefEAlphaD}
\end{equation}
The definition of the desired deviation can be derived from~\autoref{eq:RelationEcart} by canceling the lateral deviation at the implement control point $y_T=0$ for an angular deviation of $\tilde{\theta}=0$.
As such, this modeling will effectively drive the tool position to the trajectory.
For this, any classical control law can be used, such as the one proposed by \citet{Lenain2004AdaptiveAP}.

However, this simple control law possesses some limitations.
Because of its inherent simplicity, the angular deviation $\tilde\theta$ is not directly controlled.
As such, because of the lever arm effect, a tool that is a few meters behind the robot could easily reach a substantial lateral error during the initial convergence step, or at the start of a curve.
Such a behavior is clearly not wanted as, for instance, it would mean destroying crops on neighboring rows in the context of agricultural robots.
Furthermore, such a formulation does not allow for easy improvements nor convergence guarantees.
As such, we present a novel control law based on backstepping, first regulating the desired angular deviation $\tilde\theta$, itself being controlled by the steering angle $\delta$.
As such, this method can easily monitor the tool offset as it explicitely controls the angular deviation.


\subsection{Backstepping Control of the Point $T$}\label{backstepping_control_section}
In this section, we develop a backstepping control approach to directly regulate the implement position $y_T$ as a function of the steering angle $\delta$. 
The approach consists of two stages: 1) determine the robot orientation $\tilde{\theta}^d$ required to ensure an exponential convergence of the implement's lateral deviation $y_T$; and 2) adapt the robot's steering angle $\delta$ to achieve the target orientation $\tilde{\theta}^d$. 

In the first stage, we aim to find the desired orientation $\tilde{\theta}^d$ that will lead to an exponential convergence of the tool $T$ toward the trajectory.
As such, its derivative $\dot{\tilde{\theta}}$ is not directly controlled and is assumed to be measured.
Note that instead of numerically deriving the computed angle $\tilde{\theta}^d$, this variable is preferred to be measured with sensors, as unmodeled events such as slipping, skidding or actuators delays will make the numerical estimation differ from the true rotational velocity of the robot.
As such, for clarity's sake, the measured rotional velocity will be denoted as $\bar\omega$ in the following.
Therefore, in the first stage of the backstepping approach, the kinematic model of the lateral deviation $y_T$ is rewritten as
\begin{equation}
    \dot{y}_T = v\sin\tilde{\theta}+\bar\omega\left(T_s\cos\tilde{\theta} - T_y\sin\tilde{\theta}\right).
    \label{eq:ytdot}
\end{equation}
The goal of the first backstepping stage is to find a desired angular deviation $\tilde{\theta}^d$ such that the error $y_T$ exponentially decreases to zero.

In \autoref{eq:ytdot}, the derivative is expressed with respect to time.
Consequently, designing a control law based on this equation would result in a convergence characterized by a time-based metric.
In agricultural scenarios, a more interesting quantity is the distance of convergence.
Indeed, while working in the field, the main concern is to preserve crops, and thus theoretical guaranties about the convergence in distance are of great importance.
As such, using \autoref{eq:DotTheata} and \autoref{eq:ytdot}, we reformulate the kinematic of the tool lateral deviation $y_T$ with respect to curvilinear abscissa, as
 \begin{equation}
\begin{aligned}
		y'_T &= \dv{y_T}{t}\left(\dv{s}{t}\right)^{-1} \\
             &= \alpha\left[\tan\tilde\theta+ \gamma\left(T_s+T_y\tan\tilde\theta\right)\right],
\end{aligned} \label{eq:DerivYT}
\end{equation}
with
\begin{equation}
\begin{aligned}
    \alpha &= 1-c(s)y, \quad\text{and }
    \gamma =\frac{\bar\omega}{v}.
    \label{eq:defalphagamma}
    \end{aligned}
\end{equation}
Unsurprisingly, as we now analyze the convergence of the lateral error $y_T$ in terms of distance, it is necessary for the robot's velocity $v$ to be non-zero.

As the vehicle steers toward the path, the point $T$ initially moves away before gradually converging. Thus, pure exponential convergence define as $y_T'=-k_y y_T$, $k_y>0$ does not accurately describe the kinematics, requiring a specific convergence form to cancel the lateral error $y_T$:
\begin{equation}
\begin{aligned}
		y'_T & = -k_y y_T + \alpha \gamma T_s,
\end{aligned} \label{eq:Cond01}
\end{equation}
with $k_y>0$ setting the convergence distance of the implement's lateral deviation $y_T$. 
The term $\alpha \gamma T_s$ converges to zero as 
$\gamma =\frac{\bar\omega}{v}$ approaches zero, ensured by the angular error dynamics imposed at the second stage of the method described below. 
Applying this constraint on $y_T'$, the desired angular deviation $\tilde\theta^d$ is given by
\begin{equation}
\begin{aligned}
		\tilde\theta^d & = \arctan \left( \frac{-k_y \frac{y_T}{\alpha}}{1 - \gamma T_y}  \right).
\end{aligned} \label{eq:Step01}
\end{equation}
Two numerical singularities exist at $y=1/c(s)$, which is the same as in \autoref{eq:DotTheata}, and at $1-\gamma T_y=0$:
rearranging the terms, we arrive to the singularity at $v/\bar\omega = T_y$. 
Intuitively, this means that if the point to be controlled is at the same distance as the turning radius of the vehicle, no orientation can satisfy \autoref{eq:Cond01}.
Once again, such an event is unlikely in most applications, as the implement will remain close to the vehicle compared to its turning radius.

If the angular deviation of the robot is equal to the value $\tilde\theta^d$, the condition of \autoref{eq:Cond01} is satisfied and the lateral deviation of the implement converges towards the reference trajectory. 
As such, the remaining task is to guarantee a fast convergence of the angular deviation $\tilde{\theta}$ towards the desired value $\tilde\theta^d$.
For the second backstepping stage, we define the error $e_\theta=\tilde\theta-\tilde\theta^d$, and neglecting the variations of $\tilde\theta^d$, we can write the spatial derivative of this error as
\begin{equation}
\begin{aligned}
		e'_\theta &= \dv{e_\theta}{t}\left(\dv{s}{t}\right)^{-1} \\
                  &\approx \dv{\tilde\theta}{t}\left(\dv{s}{t}\right)^{-1} \\
                  &= \frac{1-c(s)y}{L\cos{\tilde\theta}}\tan\delta - c(s).
\end{aligned} \label{eq:derivEtheta}
\end{equation}
As done in the first stage, we guarantee the exponential convergence of the vehicle's orientation toward the desired value by setting the differential equation
\begin{equation}
\begin{aligned}
		e'_\theta & = -k_\theta e_\theta,
\end{aligned} \label{eq:Cond02}
\end{equation}
with $k_\theta>0$ the gain setting the convergence distance. 
Note that the derivative of $\tilde{\theta}$ converges to zero, ensuring the convergence of the first stage in \autoref{eq:Cond01}.
Thus, solving \autoref{eq:Cond02}, the steering angle that ensures the convergence of the angular deviation to $\tilde\theta_d$ is
\begin{equation}
\begin{aligned}
		\delta^d & = \arctan (L\frac{\left[-k_\theta e_\theta + c(s) \right]\cos{\tilde\theta}}{1-c(s)y}).
\end{aligned} \label{eq:ControlLawB}
\end{equation}
The angular deviation of the vehicle will converge to the desired value $\tilde\theta_d$, and subsequently the control point $T$ will exponentially converge to the trajectory.
However, one must be careful to set the gains $k_y, k_\theta$ so that the convergence of the angular error $e_\theta$ is much faster, that is $k_\theta\gg k_y$.

In conclusion, given the lateral and angular deviations $y,\tilde\theta$ measured by on-board sensors, the tool error $y_T$ can be computed using \autoref{eq:RelationEcart}. From this, the current desired angular deviation $\theta_d$ is determined with \autoref{eq:Step01}, and the steering angle to apply to the robot is finally found with \autoref{eq:ControlLawB}, driving the tool toward the trajectory.

\section{Experiments}
In this section, we provide an analysis of the proposed methods in a real world scenario.
First, we present a comparison between the desired deviation and backstepping approaches, showing that the backstepping method reaches similar to better results while proposing theoretical guarantees, as discussed earlier in this paper.
Then, we provide an analysis of the impact of the tool's position on the robot, showing a clear correlation between the path tracking error and the tool position.

\begin{figure}[thbp]
\centering
\includegraphics[width=\linewidth]{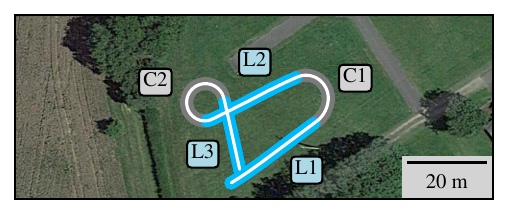}
\caption{Reference trajectory used in the experiments. It consists of three straight lines (L1, L2, L3) and two curves (C1, C2) of different curvatures.}
\label{fig:Expe_setup}
\end{figure}

\subsection{Experimental setup}
The following experiments were conducted using the robot shown in \autoref{fig:intro}.
The vehicle is a 4-wheel-drive Ackerman-type robot, able to tow numerous implements, such as mechanical weeders. 
It is equipped with a \ac{RTK} \ac{GNSS} sensor, used to compute the lateral and angular deviations $y, \tilde\theta$ from a geo-referenced, pre-recorded trajectory.
The term $\gamma = \frac{\bar\omega}{v}$ in the first stage of the backstepping method is obtained by measuring the steering angle $\delta$ with an encoder sensor, with $\bar\omega$ computed using \autoref{eq:DotTheata} and the measured steering angle. 
The experiments were conducted in an experimental farm, with the reference trajectory depicted in \autoref{fig:Expe_setup}.
It consists of three straight lines (L1, L2, L3) connected by two curves (C1, C2) of different curvatures, with opposite signs.
In order to simplify the analysis, the whole trajectory is performed on grass, thus minimizing the changes in grip conditions.
Finally, the controller gains are set to $k_y = 0.21$, $k_{\theta} = 0.63$, allowing a convergence distance of $\SI{15}{\m}$.
The velocity of the robot was set to $v=\SI{0.75}{\m\per\s}$.

\subsection{Comparison} \label{comparison}
\begin{figure*}[t]
      \centering
      \vspace{-0.25mm}
       \includegraphics[width=\linewidth]{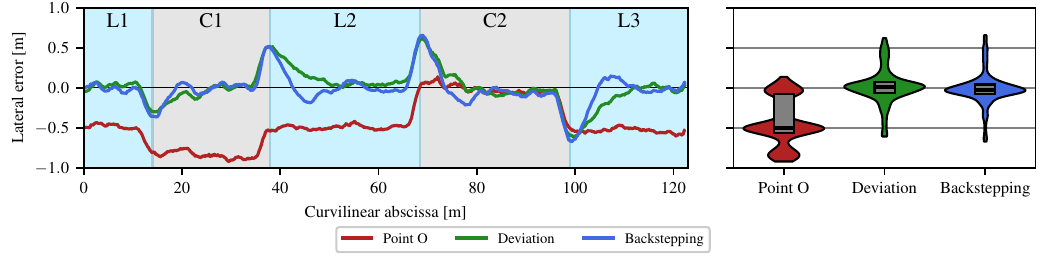}
      \caption{Implement lateral error of a classical (control of the center point $O$), desired deviation, and backstepping approaches. Left: Lateral error as the function of the curvilinear abscissa. The lines (L1, L2, L3) and curves (C1, C2) correspond to the areas highlighted in \autoref{fig:Expe_setup}. Right: Overall distribution of the errors for each method, with the median and quartiles at 25\% and 75\% represented as box plots.}
      \label{fig:compare}
\end{figure*}
We provide an analysis of the two proposed methods to control an offset point of the robot: a direct extension of the classical control methods, and a more complex one based on backstepping.
In this experiment, the offset control point was set to $T_s = \SI{-2.5}{\m}$ and $T_y =\SI{-0.5}{\m}$, which corresponds to a location behind and to the right of the robot.
For the desired deviation method, the control algorithm described in \cite{Lenain2004AdaptiveAP} is used, with the parameters set to have the same theoretical convergence distance.
\autoref{fig:compare} shows the lateral deviation error of the implement as a function of the traveled distance.
For reference, the error of the implement while using a classical control algorithm of the rear axle (point $O$ in \autoref{fig:model}) is provided.
Inevitably, the classical approach of controlling the rear axle center lead to a great amount of error:
the median error corresponds to the lateral shift of the implement, that is in our case $T_y =\SI{-0.5}{\m}$.
Furthermore, the lever arm effect increases the error while turning left (curve C1) and decreases it while turning right (curve C2).
As such, this clearly motivates the need for implement-aware control laws. 
Indeed, even if the lateral shift of the tool is null, \ie $T_y=0$, the lever-arm effect is not taken into account, thus leading to great errors in curves \cite{gan2007implement}.
Then, the first proposed method is to shift the desired lateral deviation to make the implement follow the trajectory.
As seen in \autoref{fig:compare}, this simple extension provides reliable results, as the tool is effectively driven toward the trajectory.
However, one can note a longer convergence distance at the end of the first curve, compared to the backstepping method.
Indeed, in the desired deviation method, the control relies on a direct shift of the lateral error, with a desired angular error of always zero.
The backstepping method is able to converge quicker by setting a nonzero desired angular deviation, thus recovering in a shorter time at the end of the curve.
The effect is lessened in the next curve (C2), as the implement is moved in the right direction because of the level arm effect during the turn.

\subsection{Impact of the Tool Position}
To assess the effect of the offset point $T$ position on the control law precision, we conducted experiments while varying the relative position of the point $T$, on the same trajectory and constant longitudinal speed.
The lateral distance $T_y$ was fixed at $\SI{-0.5}{\m}$, with the longitudinal position $T_s$ varying between $\SI{-2}{\m}$ and $\SI{2}{\m}$, as being the parameter that induces the greatest variation of error in the experiments.
The backstepping method was used in this experiment.

\begin{figure}[t]
      \centering
      \vspace{-4mm}
       \includegraphics[clip,width=\linewidth]{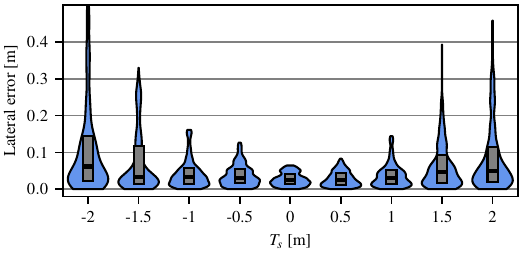}
      \caption{Distributions of the absolute deviation error as a function of the implement's longitudinal offset $T_s$. The box plots represent the medians with quartiles at 25\% and 75\%.}
      \label{fig:impactTs}
      \vspace{-4mm}
\end{figure}

~\autoref{fig:impactTs} depicts the absolute lateral error of the implement for different values of the longitudinal offset $T_s$. 
With no surprise, the lateral error is greater when the implement is farther from the robot's center, as the lever arm effect becomes more important.
Furthermore, one can see that the error is bigger when the implement is at the rear ($T_s < 0$) of the vehicle compared to the same distance at the front:
we theorize this phenomenon is attributed to the anticipatory nature of the control when the implement is located at the front of the vehicle, thus lessening the implement's lateral offset. 
In the absence of predictive control, these errors are amplified as the system reacts to errors after they occur rather than anticipating and adjusting for them proactively. 
As such, a predictive control approach could address this issue by predicting future states of the system, mitigating the effect of the lever arm.

\section{Conclusion}
In this paper, we proposed a novel method for controlling an offset point rigidly linked to the robot. 
The tracking of an offset point on an implement is a critical requirement in agricultural robotics, as the implement is responsible for performing the agronomic tasks. 
We showed that control methods focused solely on the vehicle’s center are inadequate for performing agricultural tasks, particularly in the case of large implements. 
To address this limitation, we propose two alternative approaches.
The first adapts classical control methods to manage the desired lateral deviation of the implement, while the second approach employs backstepping techniques to compute an angular deviation that ensures an exponential convergence of the lateral error. Both methods successfully converge the implement’s control point to the reference trajectory, while the backstepping approach offers stronger theoretical guarantees and achieves shorter convergence distances.

Future work will explore the integration of feedforward and predictive control strategies to mitigate the lever arm effect and enhance the overall precision. 
Additionally, we will investigate the simultaneous control of multiple points on the robot, as it is critical not only to manage the implement position but also to ensure that the robot’s wheels avoid damaging crops.






\section*{ACKNOWLEDGMENT}
This work has been funded by the french National Research Agency (ANR), under the grant ANR-19-LCV2-0011, attributed to the joint laboratory Tiara (\url{www6.inrae.fr/tiara}). It has also received the support of the French government research program "Investissements d'Avenir" through the IDEX-ISITE initiative 16-IDEX-0001 (CAP 20-25), the IMobS3 Laboratory of Excellence (ANR-10-LABX-16-01).  This work has been partially supported by ROBOTEX 2.0 (Grants ROBOTEX ANR-10-EQPX-44-01 and TIRREX ANR-21-ESRE-0015)


\printbibliography

\end{document}